\documentclass[superscriptaddress,pra,showpacs,twocolumn]{revtex4} 
\usepackage{graphicx}
\usepackage{float}
\usepackage{epstopdf}
\usepackage{color}
\usepackage{amsmath}
\usepackage{amssymb}

\begin{document}
	
\title{Quantum regime for the nuclear energy loss of fast atoms above crystal surfaces}
\author{P. Roncin and M. Debiossac}

\affiliation{Institut des Sciences Mol\'{e}culaires d'Orsay (ISMO), 
	CNRS, Univ. Paris-Sud, Universit\'{e} Paris-Saclay, F-91405 Orsay, France}
\date{\today}


\begin{abstract}
	 To describe the grazing scattering of keV atoms at surface, a new quantum binary collision model have been proposed where the dynamical properties of the surface atoms are considered via the wave-function of the local Debye harmonic oscillator. This leads to a finite probability of elastic scattering where the momentum transferred during the successive binary collisions is not associated with a change of energy. This Lamb-Dicke regime of the multiple collisions at the surface produces the same coherence ratio as the modified Debye-Waller factor adapted to grazing angle fast atom diffraction (GIFAD) but with the additional ability to predict the spot shape of the inelastic diffraction profiles. In terms of energy loss, we show here that at low angle of incidence $\theta$,  this Lamb-Dicke effect leads to a marked $\theta^7$ dependency progressively merging to the $\theta^3$ classical dependency. The analytic model presented is supported by numerical simulations for neon atoms scattered off a LiF surface and remains to be confirmed by experiment.
\end{abstract}

\maketitle

\section{Introduction}

Several techniques have develop to probe mater by keV ions or atoms.
The interpretation often relies on quasi head-on binary collisions to identify the mass of the encountered atom.
The mean penetration depth is adjusted by tuning the energy and incidence angle while the outgoing angles and energy are analyzed to unravel the kinematics and estimate the mean path inside the solid~\cite{Bauer_2007} or liquid~\cite{Andersson_2005}. 
The numerous applications range from material analysis, radiation hardening to the design of detectors for high energy physics.
Theoretically, one considers that the energy transfer to the electrons and to the nuclei are separate contributions and numerous simulation programs have been developed.

The situation is different above surfaces where the importance of close collisions can be reduced under grazing incidence geometry.
On metals, several electron capture and loss processes were identified and referred to $Z_1$ oscillations. These oscillations origin from atomic shell effect, see e.g.~\cite{Winter_rep_2002} for a review.
With LiF, a large band gap insulator, and measuring projectile energy loss in coincidences with emitted electrons \cite{Roncin_98,Roncin_2002}, individual quasi-molecular electronic processes were identified; transient negative ion formation, electron detachment, surface excitons population, bridging the gap with perfectly defined gas phases collisions.
At lower incidence angle, when no electronic excitation of the surface is observed, significant energy loss could be attributed to the sole excitation of optical phonons by the electric field of the fast ionic projectile~\cite{Borisov_99,Villette_2000} or, reversely resonant electronic excitation of the projectile by the periodic field of the surface \cite{Auth_1997}.
For keV neon atomic projectile, a nuclear energy loss in the eV region could be measured at grazing incidence \cite{Mertens_2000} and reproduced by classical trajectory simulations.

Reducing further the angle of incidence, the diffraction of the fast ($\sim$keV) projectile could be observed \cite{Rousseau_2007,Schuller_2007}, allowing precise measurement of the surface topology \cite{DebiossacPRB_2014} in strong analogy with low energy ($\sim$50\:meV) atomic diffraction \cite{FariasCPL2004}.
It also outlines that the successive binary collision reflecting the projectile few \AA\: above the surface can be gentle enough to preserve its coherence.
For grazing enough angles $\theta$, these collisions enter the Lamb-Dicke regime of so called recoilless emission. 
The interpretation is the following: when the recoil energy of the surface atoms becomes less than the Debye energy $\hbar\omega_D$, the probability of vibrational excitation vanishes, allowing purely elastic diffraction of the projectile as in the perfectly periodic surfaces considered in theoretical descriptions \cite{Zugarramurdi_2012,Sanz,Diaz_2016b,Aigner_2008,Gravielle_2014}.
In experiments, both elastic and inelastic diffraction have been observed \cite{Busch_2012,DebiossacPRB_2014}. The quantum binary collision model \cite{Roncin_2017} was developed to describe the individual collisions in this regime. It is applied here to suggest a marked $\theta^7$ dependence of the mean projectile energy loss. The model is briefly recalled starting from classical mechanics.

\section{Classical binary collision model}
In the gas phase, a binary collision between a projectile having an energy $E_0$ and a target at rest is called elastic when no electronic excitation takes place. 
The conservation of energy and momentum connects the scattering angle and the projectile energy loss so that for small scattering angle, $\theta\sim\sin\theta$ the target recoil momentum $\vec{\delta p}$ is almost perpendicular to the projectile and the recoil energy is $E_r = \mu E_0 \theta^2$ where $\mu=m_p/m_c$ is the ratio of the projectile to target mass. 

On surfaces, the specular reflection of a fast atom corresponds to the exchange of a momentum $2 k_0 \sin\theta$ with the surface ($k_0=\sqrt{2m_p E_0}$). 
However, since grazing incidence correspond to multiple successive shallow collisions with the surface atoms, much less momentum is exchanged in each binary collision. This can be turned quantitative by the planar approximation.
Assuming that the interaction takes place few \AA\: above the surface, the Schr\"{o}dinger equation indicates that the valence electron wave function $\phi(x,y,z)$ should behave as $e^{-\Gamma z}$ with $\Gamma \sim \sqrt{(2W)}$ where $W$ is the height of the barrier that the electron sees when trying to escape to vacuum, \textit{i.e.} the surface work function. 
Neglecting the van der Waals forces, a noble gas projectile trying to penetrate the surface is repelled by the Pauli repulsion proportional to the surface electronic density $\rho(x,y,z) \propto |\phi\rangle|^2$.
Neglecting for a moment the weak surface corrugation, the planar average interaction potential $V_{1D}(z)$ =$\int\int V_{3D}(x,y,z)dxdy$ should also display the same exponential character $V_{1D}(z)\propto e^{-\Gamma z}$ few \AA\: above the surface.
On such a planar potential the projectile trajectory $z(t)$ is analytic and so are the first and second derivative representing the velocity and acceleration (see e.g.~\cite{Manson2008,Roncin_2017}).
This acceleration curve $\ddot z(x)$ describes the density of momentum exchange with the surface and is quasi-gaussian with a fwhm $L=\alpha /(\Gamma \theta)$ with $\alpha=4\cosh^{-1}(2^{1/4})\sim 2.42$ which provides a clear definition of the trajectory length $L$.
The $1/\theta$ dependence outlines the rapid increase of the trajectory length at grazing incidence and was noted in trajectory simulation \cite{Villette_these}.

To transform this projectile acceleration into a recoil momentum $\vec{\delta p}$ of each surface atom, the binary collision approximation is used. 
It assumes that only one surface atom at a time contributes to the acceleration and that the linear density $n_x$ is, for instance, one per lattice unit $n_x = 1/a$. 
The classical energy loss $\Delta E_{Cl}$ is the sum of these recoil energies $E_r = \vec{\delta p}^2 /2m_c$ along the surface trajectory  \cite{Manson2008,Roncin_2017}.
\begin{equation}
\Delta E_{Cl} = \frac{2}{3} \mu E \Gamma a \theta_{in}^3  \label{Eloss}        
\end{equation}
Compared with the individual binary recoil energy loss that would originate from a single collision, $\Delta E_{Cl}$ is $N_{eq}$ times smaller with $N_{eq}$ given by 
\begin{equation}
N_{eq}= \frac{6}{\Gamma a \theta_{in}}    \label{Neq}        
\end{equation}
This number $N_{eq}$ is a reformulation of the trajectory length defined above ($N_{eq} \sim L/a$) and can be interpreted as the number of binary collision contributing equally to a small deflection $\delta \theta=2\theta_{in}/N_{eq}$. 
As the angle of incidence decreases, a smaller amount of energy loss is shared among an increasing number of scatterers so that each one undergoes a classical recoil energy scaling with $\theta ^4$:
 
\begin{equation}
E_r = \mu E\delta \theta^2 \simeq \frac{\Gamma^2 a^2}{9} \mu E \theta_{in}^{~4}   \label{recoil}        
\end{equation}

where the second member is simply the overall classical energy loss in Eq.\ref{Eloss} divided by the effective number of collider $N_{eq}$ in Eq.\ref{Neq}.
This planar model is a uniform mirror so that the angular distribution is delta functions, $\theta_{out} = \theta_{in}$ and $\phi_{out} = 0$.
Changing to a more realistic egg-carton potential energy surface with atoms at equilibrium position and scanning different impact parameters also produces very narrow scattering functions located on the Laue circle ($\theta_{out} = \theta_{in}$) provided the angle of incidence is small enough.
This situation where only the impact parameter perpendicular to the beam direction is relevant corresponds to the axial channeling approximation initiated in classical scattering~\cite{danailov} and extended to elastic diffraction \cite{Zugarramurdi_2012}.

The thermal agitation breaks this axial symmetry and and a broad angular scattering profile is observed well fitted by a log-normal distribution \cite{Pfandzelter_1998,Villette_2000,Villette_these}. 
\begin{equation} LN[\theta_0;w](\theta)=\frac{A}{\sqrt{2\pi}w\theta}exp(\frac{-(\ln \frac{\theta}{\theta_0})^2}{2w^2} )  \label{LN} \end{equation}
Where $\theta_0$ is the median scattering angle, $w$ is a relative width and the variance is $\sigma^2_\theta=e^{w^2}(e^{w^2}-1)\theta_0^2\label{sigma_LN}$.
This asymmetric scattering distribution was found to originate from the individual binary collisions~\cite{Manson2008,Roncin_2017}.
Let us consider a screened coulombic binary interaction potential  $V(R)=\frac{A}{R} e^{-\Gamma R}$ which naturally integrates to the planar form $V_{1D}(z)=A/\Gamma e^{-\Gamma z}$ used above. Then, the scattering angle resulting from a single collision at an impact parameter $z=b$ scales as $\delta \theta\propto e^{-\Gamma b}$.
The thermal fluctuation of the surface atom by a displacement $dz$ gives rise to a log-normal deflection profile of relative width $w=\Gamma \sigma_z$ independent of the exact value of the impact parameter $b$ or of the individual deflection angle~\cite{Manson2008,Roncin_2017}.

At a temperature $T$, the standard deviation $\sigma_z$ derived from the quantum position distribution taking into account the ground state motion at zero temperature is 
\begin{equation}  \sigma_z^2 = \langle   z^2  \rangle  =\frac{3\hbar^2}{2m~k_B T_D}\coth(\frac{T_D}{2T}) \label{z2T} \end{equation}
where $T_D$ is the Debye surface temperature describing the local harmonic oscillator, $k_B$ is the Boltzmann constant so that $\hbar\omega_D=k_B T_D$ is the energy of a vibration quantum of the Debye oscillator. 

This allows a direct connection between the individual position fluctuations $\sigma_z$, the deflection fluctuations $\sigma_\theta$ and the recoil energy value $E_r$ ; $\sigma^2_\theta=e^{w^2}(e^{w^2}-1)\delta\theta^2 = \beta E_r$ with $\beta=\frac{e^{w^2}(e^{w^2}-1)}{\mu E}$ and $E_r=\mu E \delta\theta^2$. 

The overall scattering profile resulting from successive convolutions of such log-normal profiles have a variance equal to the sum of the individual variances: $\sigma^2_{tot} =\Sigma \sigma^2_\theta$ directly related to the total energy loss
\begin{equation} \sigma^2_{tot} =e^{w^2}(e^{w^2}-1)\Sigma\delta\theta^2~=~\beta\Sigma E_r~=~\beta \Delta E_{Cl} \label{sigma_tot} \end{equation}

Coming back to the energy loss of a single collision, a position fluctuation by $dz$ produces a modified recoil energy $E_r'=\mu E~\delta\theta'^2 = E_r e^{-2\Gamma dz}$.
This indicates that $P_{Cl}(\Delta E)$ is also a log-normal distribution with a median value $\Delta E_{Cl}$ given by Eq.\ref{Eloss} and a relative width $w'=2w=2\Gamma\sigma_z$ twice larger than that of the angular profile due to the quadratic dependence to a position fluctuation.
\begin{equation} P_{Cl}(\Delta E)=LN[\Delta E_{Cl};2\Gamma\sigma_z](\Delta E)  \label{LNE} \end{equation}
The mean value of this asymmetric profile is given by
\begin{equation} \langle\Delta E \rangle = \Delta E_{Cl}~ e^{2\Gamma^2\sigma_z^2}  \label{MeanDE} \end{equation}
which converges to $\Delta E_{Cl}$ only for zero vibration amplitude which never occurs even at zero temperature (Eq.\ref{z2T}).
In this approach the overall scattering and energy loss profiles have been derived with a rigid lattice surface where the atoms are sitting at equilibrium positions but taking into account classically the broadening of the scattering distributions induced by the positions fluctuations. 


\section{The quantum binary collision model: QBCM}
The quantum binary collision model introduces the dynamic properties of the harmonic oscillator by its wave function $|\psi\rangle$. 
Since space and momentum can not be treated independently, the collision is considered with the wave-function of the surface atom. It means that in  Eq.\ref{z2T} $\langle   z^2  \rangle $ only describes the spatial extension of the wave-function centered at the lattice equilibrium position.
 
In-line with the sudden approximation used above, the probability for a transition between an initial state $|\psi_i\rangle$ and a final state $|\psi_f\rangle$ in response to a momentum exchange  $\vec{\delta p}$ is given by $p_{if} = |\langle\psi_f| e^{i\delta p} |\psi_i\rangle|^2$. 
In the present context, we are primarily interested to the probability $p_e$ that the collision is elastic, i.e. proceeds without exchange of energy. 
This requires that no vibration excitation takes place and corresponds to $p_e= |\langle\psi| e^{i\delta p} |\psi\rangle|^2$. 
This probability $p_e$ describes the ability of the wave function to absorb the momentum $\vec{\delta p}$ while remaining unchanged and can be significant only if $\vec{\delta p}$ is present in the wave function. 
It can be evaluated as $p_e = e^{\delta p^2 \langle z^2 \rangle}$ with help of the Bloch theorem \cite{Cohen_Tannoudji}.
For the ground state it amounts to $p_e = e^\frac{-E_r}{\hbar\omega}$. At variance with the classical interpretation, $E_r=\delta p^2 /2m$ is here the recoil energy that is not actually exchanged in the process.
This corresponds to the Lamb-Dicke probability of recoilless emission well known in spectroscopy and in cold atom optical lattices.
It also corresponds to the Debye-Waller factor, ubiquitous in crystallography and usually derived as the coherence ratio $e^{-\delta \phi^2}$ of a Gaussian phase distribution. The phase $\delta \phi \sim 2 k \sigma_z$ corresponds to the scattering of particle with a wavenumber $k$ reflected by an ensemble of atoms normally dispersed by $\sigma_z$ around their equilibrium positions.
Using the thermally averaged value of $\langle z^2 \rangle$ in Eq.\ref{z2T} the value of $p_e$ reads
\begin{equation} p_e= exp(-\frac{3 E_r}{k_B T_D} \coth(\frac{T_D}{2T}) )  \label{pe} \end{equation}
The Lamb-Dicke effect becomes significant when the classical recoil energy $E_r$ is less than  the discrete energy spacing $\hbar \omega_D = k T_D$. 
Depending on the exact projectile energy E and mass, this will eventually occur around few degree because of the very rapid fourth power dependence $E_r \propto\theta^4$ in Eq.\ref{recoil}. 
At 1 deg, $\theta^4\sim 10^{-7}$ transforming few keV projectile energy into meV recoil energy.

The probability that all the successive binary collisions along a complete trajectory proceed in the elastic regime is given by the product probability $P_e=\Pi ~p_e$.
It factorizes with the sum of the individual recoil energies in the exponent putting forward the overall energy loss $\Delta E_{Cl}=\Sigma E_r$ in Eq.\ref{Eloss}. Note that this result remains unchanged for all phonon modes constructed on these momentum transfer because the dispersion curve of the transverse optical mode is flat \cite{Bilz_79}.
\begin{equation} P_e= exp(\frac{-3 \Delta E_{Cl}}{k_B T_D} \coth(\frac{T_D}{2T}) ) \label{Pes} \end{equation}

\section{The quasi-elastic regime}

The quasi-elastic regime can be defined by $P_e\sim 1$ and the complementary probability that at least one binary collision is inelastic $\bar{P}_e=1-P_e$ can be expanded in Taylor series with a leading term (Eq.\ref{Pebar}). Assuming that, in this regime, a single or few inelastic events take place with a mean recoil energy $ E_r$ from Eq.\ref{recoil}, the overall energy loss should scale as $\Delta E_{Qu} \sim \bar{P}_e E_r$ (Eq.\ref{DEbar}).

\begin{equation} \bar{P}_e= (1-P_e)~\sim \frac{3 \Delta E_{Cl}}{k_B T_D} \coth(\frac{T_D}{2T})  \label{Pebar} \end{equation}
\begin{equation} \Delta E_{Qu}~\sim \frac{2 }{9 k_B T_D}    \coth(\frac{T_D}{2T}) \mu^2 E^2 \Gamma^3 a^3 \theta_{in}^7 \label{DEbar} \end{equation}
\begin{equation}\Delta E_{Qu} = \Sigma_{n=0}^{n=N_{eq}}~n~E_r~\binom{N_{eq}}{n}~\bar{p}_e^{~n}~p_e^{N_{eq}-n} \label{expension_1} \end{equation}

Alternately, using the equivalent scatterer model (Eq.\ref{Neq}) the Taylor expansion can be applied to the individual probability in Eq.\ref{pe}, $\bar{p}_e=1-p_e\sim \frac{3 E_r}{k_B T_D} \coth(\frac{T_D}{2T})$ allowing an expansion in terms of the number $n$ of inelastic events (Eq.\ref{expension_1}). The result of Eq.\ref{DEbar} is identical to the first non zero term in the expansion : $\Delta E_{Qu} = N_{eq}~E_r~\bar{p}_e~p_e^{N_{eq}-1}$ giving $\Delta E_{Qu} = ~\frac{3}{k_B T_D} N_{eq} E_r^2\coth(\frac{T_D}{2T})$.

\section{3D Trajectory simulations}
Beyond the perturbative approach the individual Lamb-Dicke probabilities have been calculated  by integrating numerically the projectile trajectory  on the rigid lattice with surface atoms at equilibrium position.
At any moment, the projectile interacts with all surface atoms within a sphere radius of 20 \AA\:  and the momentum $\vec{dp_i}=\vec\gamma_i dt$ exchanged with each surface atom is tracked at each time step.
At the end of the trajectory, the exact classical recoil momentum $\vec{p_i}=\int_{-\infty}^{\infty} \vec{dp_i}~dt$ and recoil energy $E_i=|\vec{p_i}|^2/2m_i$ are calculated for each surface atom. 
The temperature dependent mean energy loss is then the weighted sum $\Delta E_{Qu} = \Sigma_i E_i~\bar{p_e} (E_i)$. 
 
\begin{figure}	\includegraphics[width=0.85\linewidth]{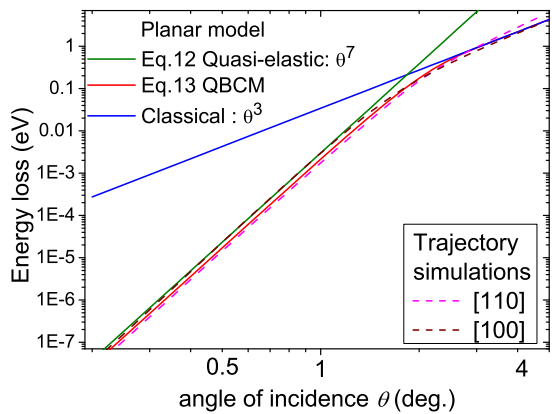}
	\caption{Energy loss of 1 keV neon projectiles scattered off a LiF surface calculated with the planar model and trajectory simulations with $\Gamma =3.2 \AA^{-1}$ and $T_D =340K$. Both show a marked $\theta^7$ dependence below two deg. incidence when the Lamb-Dicke effect is taken into account.
\label{fig_lnEloss}}\end{figure}

\begin{figure}	\includegraphics[width=0.85\linewidth]{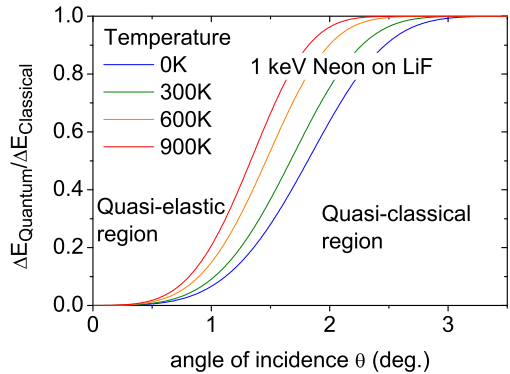}
	\caption{Ratio of the quantum $\Delta E_{Qu}$ (Eq. \ref{expension_1}) to classical energy loss $\Delta E_{Cl}$ (Eq. \ref{Eloss}). It indicates where the Lamb-Dicke effect significantly reduces the projectile energy loss.\label{fig_Eloss_ratio}}\end{figure}

This approach gets rid of the binary collision assumption and the results can be averaged over all impact parameters within the lattice unit so that different crystallographic directions can be probed without restriction on the nature and position of the atoms in the lattice unit. For instance the lithium atoms neglected in the planar model can be accounted for. 
The energy loss calculated  for 1keV neon projectiles impinging on a LiF surface along the $[110]$ and $[100]$ directions are reported on Fig.\ref{fig_lnEloss} together with the analytic planar model in its classical (Eq.\ref{Eloss}) and quantum (Eq.\ref{expension_1}) forms. 
The general $\theta^7$ dependence is very well reproduced by all models taking into account the Lamb-Dicke effect.
To illustrate the transition from the quasi-elastic to the classical regime, we plot in Fig.\ref{fig_Eloss_ratio} the ratio of the  quasi-elastic to the classical energy loss at different temperatures.  
It provides a simple illustration of the Lamb-Dicke effect considering the quantum nature of the surface. 
This ratio is almost zero in the quasi-elastic regime with an asymptotic form at low incidence ;
\begin{equation} \frac{\Delta E_{Quantum}}{\Delta E_{Classical}}= \frac{\mu E \Gamma^2 a^2}{3~k_B T_D}  \coth(\frac{T_D}{2T})  \theta^4~\label{ratio} \end{equation}
The ratio has intermediate values in the mixed regime where both elastic and inelastic values become significant before reaching unity in the quasi-classical regime.
The magnitude of the energy loss is rather small making experiments difficult, however the model indicates a direct correspondence of the energy loss with the angular straggling and the elastic fraction.
Provided the surface quality is large enough to reduce the contribution of defects to a negligible value, this energy scaling could be observed in the angular domain. 

Another output of the present simulations is that, in the quasi-elastic regime the energy loss is comparable along the $[100]$ and $[110]$ directions as visible on Fig.\ref{fig_lnEloss}. 

Assuming as suggested above that angular straggling and energy loss are connected, this is consistent with the observation by Seifert \textit{et al.}~\cite{Seifert_2015} that, in the diffraction regime, the transverse line broadening is independent of the crystal orientation. 
At larger angles of incidence, the simulations indicate that the energy loss along both directions start to show significant differences.

\section{Summary and Conclusion}
The energy loss of grazingly scattered keV atoms on top of a crystalline surface is investigated using a simplified Lamb-Dicke formulation of the energy exchange to the phonon system. 
Neglecting the attractive contribution to the interaction potential, an analytic expression have been derived in the binary collision approximation. The quasi elastic regime previously identified in the angular scattering profile \cite{Roncin_2017} is found to be associated with a characteristic $\theta^7$ scaling before merging to a classical $\theta^3$ dependence. 
This prediction of a drastic reduction of the nuclear energy loss, if confirmed by measurement, could open a window to identify other small contributions to the energy loss such as the creations of electron-hole pair at the Fermi edge \cite{Bunermann_2015}. 
In general, the proper account of quantum effect leading to decoherence should allow fast atom diffraction to provide more quantitative information on the surface promoting application such as in-situ growth monitoring in molecular beam epitaxy \cite{Atkinson_2014,Debiossac_ASS_2017}.


\end{document}